\documentclass[12pt, final]{article}
\usepackage[utf8]{inputenc}
\usepackage{amsmath}
\usepackage{mathtools}
\usepackage{amsbsy}
\usepackage[utf8]{inputenc}
\usepackage{graphicx}
\graphicspath{{images/}}
\usepackage[a4paper,margin=1in]{geometry}
\usepackage{fancyhdr}
\usepackage[linesnumbered,ruled,vlined]{algorithm2e}
\usepackage{bbm}
\usepackage{subcaption}
\usepackage{float}
\usepackage{multicol}
\usepackage{threeparttablex}
\usepackage{amsfonts}
\usepackage{listings}
\usepackage{apacite}
\usepackage{enumitem}
\usepackage{natbib}
\usepackage{booktabs}
\usepackage[table, svgnames, dvipsnames]{xcolor}
\usepackage{makecell, cellspace, caption}
\usepackage{siunitx}
\usepackage{natbib}
\usepackage{multirow}
\usepackage{setspace}
\usepackage{lipsum}
\makeatletter
\renewcommand{\maketitle}{\bgroup\setlength{\parindent}{0pt}
\begin{flushleft}
  \textbf{\@title}

  \@author
\end{flushleft}\egroup
}
\makeatother

\title{\Large Bayesian design for mathematical models of fruit growth based on misspecified prior information \\
  ~ \\
}

\author{
  \normalsize Najimuddin, N.$^{*1,2}$, Warne, D.J.$^{1,2}$, Thompson, H.$^{1,2}$, McGree, J.M.$^{1,2}$ \\
  \vspace{0.5cm} 
  \small $^{1}$School of Mathematical Sciences, Queensland University of Technology, Brisbane, Queensland, Australia\\
  $^{2}$Centre for Data Science, Queensland University of Technology, Brisbane, Queensland, Australia\\
  \vspace{0.5cm} 
  {$^{*}$fathimanushrath.najimuddin@hdr.qut.edu.au}
}

\begin{document}

\maketitle
\begin{abstract}

\noindent
Bayesian design can be used for efficient data collection over time when the process can be described by the solution to an ordinary differential equation (ODE).
Typically, Bayesian designs in such settings are obtained by maximising the expected value of a utility function that is derived from the joint probability distribution of the parameters and the response, given prior information about an appropriate ODE. 
However, in practice, appropriately defining such information \textit{a priori} can be difficult due to incomplete knowledge about the mechanisms that govern how the process evolves over time.
In this paper, we propose a method for finding Bayesian designs based on a flexible class of ODEs.
Specifically, we consider the inclusion of spline terms into ODEs to provide flexibility in modelling how the process changes over time.
We then propose to leverage this flexibility to form designs that are efficient even when the prior information is misspecified.
Our approach is motivated by a sampling problem in agriculture where the goal is to provide a better understanding of fruit growth where prior information is based on studies conducted overseas, and therefore is potentially misspecified.
\end{abstract}

\section{Introduction}

Data collection plays a critical role in understanding complex natural phenomena such as climate change, species extinction, ocean acidification, and infectious disease outbreaks \citep{bindoff2019changing,seebens2017no,tilman2011global,herrero2010smart,nordo2017comparative}.
A typical aim of such data collection is that it should provide insight into the underlying/governing relationships while also being efficient through the use of minimal resources. 
The challenge in achieving this aim is that we often have limited understanding of natural processes, so the design needs to remain informative and efficient despite potentially misspecified prior information about, for example, the underlying data generating process.
In this paper, we propose to address this challenge by leveraging the benefits of flexible modelling approaches when forming designs.
That is, instead of assuming a specific form for the model, an infinite class of models is assumed, and the flexibility of this modelling class is leveraged to form robust designs.
This approach is motivated by a sampling problem from an Australian avocado producer who was interested in increasing their understanding of fruit growth, development and maturity, noting that no such study had been published on avocado growth in Australian conditions.
Robust designs are needed for this sampling as the only available prior information in the literature is related research conducted in New Zealand \citep*{clark2003dry,dixon2006patterns} and South Africa \citep{moore1995effect}, and therefore may not be directly applicable.
 
In this paper, we consider a Bayesian framework to find designs that are robust to the misspecification of prior information. 
In terms of finding designs that are robust to model misspecification, a number of approaches have been proposed.  
For example, the most widely used approach is to average a utility function across a finite set of assumed plausible models \citep{ryan2016review,cook1982model,kristoffersen2020model,li2000model,selvaratnam2021model} which is known as the Läuter's criterion \citep{lauter1974experimental}.
This approach operates under the assumption that one of the models within the set is appropriate for describing the data, which is referred to as the $M$-closed perspective.
Even though this approach provides robustness to the form of the model, it still requires one to correctly specify the data generating process \textit{a priori}.
\citet*{woods2006designs} proposed a technique for finding robust $D$-optimal designs that accounts for uncertainty in the link function or linear predictor for generalised linear models (GLMs) with multiple explanatory variables.
Similarly, \citet*{wiens2008robust} proposed the development of robust $Q$-optimal designs for a potentially misspecified nonlinear model.
This was extended by \citet*{adewale2009robust} by considering possible misspecification of fixed effect terms in the linear predictor for logistic regression models. 
Of note, the main limitation of these approaches is that we need to specify a plausible underlying data generating model. 
However, in practical applications, defining such a model can be difficult.
To address this, we propose to develop designs under the $M$-open perspective, which allows for the consideration of models outside a candidate set.
In doing so, we propose methods to find designs that remain efficient without requiring the data generating model to be explicitly specified \textit{a priori}. 

Throughout this paper, we consider robust Bayesian design methods for Ordinary Differential Equation (ODE) models.
Such models have been considered for design previously \citep{atkinson1993optimum,atkinson2002compound,atkinson2007optimum,rodriguez2014design,warne2017optimal,overstall2019bayesian}.
However, throughout these approaches, a single ODE model was assumed to describe the data, with no consideration that such prior information may be  misspecified in some way.
We address this limitation in this paper by allowing flexibility in these models such that they can describe a wide range of relationships over time.
Many approaches are available to allow a model to be more flexible, depending on the type of model and the data. 
In particular, spline models have been shown to provide a flexible and smooth representation of the data, thus ensuring reasonable predictive performance \citep*{suk2019nonlinear}.
\citet{de2022model} have considered splines to allow flexibility to GLMs.
Given this, we suggest that such terms may be useful in ODEs for accounting for a misspecified prior due to, for example, the influence of unmeasured external factors such as the environment and/or the misspecification of the data generating process.
Accordingly, extensions to ODE models are considered through the inclusion of spline terms.  
The subsequent model is then considered as a basis to form robust designs.

The structure of the paper is arranged in the following manner.
Section 2 provides the context and motivation for the methods proposed in this paper.
Section 3 provides the statistical modelling framework for fruit growth based on ODEs and background information about statistical inference methods for ODEs.
Section 4 describes Bayesian optimal design and an overview of our proposed approach to find robust Bayesian designs. 
Section 5 applies our proposed approach to design real sampling problems in agriculture with the goal of providing a better understanding of fruit growth and development within Australian conditions.
Finally, the paper concludes with Section 6 where we discuss key outcomes of our research and suggest possible avenues for future research.

\section{Motivation}
With the demand for fruit and other consumables continuing to increase globally, there is a commercial interest and opportunity to increase revenue from exported avocados.
The challenge in doing so is addressing this opportunity with typically limited resources (such as land availability) through making better operational and on-farm decisions to increase fruit quality and minimise post harvest losses \citep*{wang2017postharvest}.
Currently in Australia there are no sampling strategies available to provide a better understanding of avocado production, and therefore there is currently a lack of information available to improve production and operations, and ultimately improve quality of fruit for domestic and international markets.
To address this, we propose an approach to derive sampling strategies to understand avocado growth in Australia.

Proper ripening and ideal eating quality of fruit are contingent upon reaching a minimum level of maturity. 
Dry matter, which associates closely with oil content, serves as an indicator of avocado quality \citep*{lee1983maturity}.
Also, the profitability of avocados relies heavily on achieving consistent high yields from large fruit. 
Research indicates that, typically, larger fruits tend to receive higher flavor ratings compared to smaller fruits when evaluated at the beginning of the season when the market acceptability is at its minimum \citep*{soule1955relation}.
Thus, the responses that are of interest to producers and therefore the focus of this work, are fruit dry matter and fruit weight. 
Measurement of fruit growth based on fruit dry matter is destructive, whereas fruit weight can be measured non-destructively by performing repeated measurements of fruit while it is still on the tree \citep*{magwaza2015review}.
Given this, an additional consideration for understanding and developing designs for fruit weight is the  within and between fruit variability.
Such considerations inform the development of statistical models to describe fruit growth, and these are defined in the next section.

During the avocado fruit growth, there is a lag phase lasting about 10 weeks after full bloom \citep*{valmayor1967cellular}.
A rapid growth phase continues for approximately 30 weeks following full flowering, varying based on the type of cultivar and the environmental conditions. This is followed by a phase of maturity when growth decreases.
The fruit typically reach maturity, determined by the  percentage of the dry matter, in about six to eight months \citep*{lee1983maturity}.
Therefore, we consider the rapid growth period (30 weeks) as the sampling period to assess dry matter.
However the fruits can be \lq tree stored\rq\ as long as 12 to 18 months after flowering in some environments \citep*{yahia2011avocado}.
Therefore, we consider 350 days as the sampling period to assess fruit weight.

Previous research into assessing fruit quality based on indicators of growth, such as fruit mass, diameter or dry matter has shown that typically such indicators follow a sigmoidal growth curve \citep*{valmayor1967cellular}.
Avocado dry matter accumulation typically follows a sigmoidal curve during its development as described by earlier researchers \citep*{daley2024preharvest,bayram2019determination}.
Some fruits, even from the same family, may exhibit a single sigmoid pattern, while other fruits can exhibit a double sigmoid pattern \citep*{pei2020comparative}. 
In some cases, especially with certain avocado cultivars or under specific growing conditions, avocado weight can exhibit a double sigmoidal curve.
This means that the two growth periods are separated by a temporary slowdown or plateau typically in winter, with the second phase continuing until harvest \citep*{pak2002fruit}. 
Factors such as climate, soil conditions, irrigation, and pruning practices can influence the shape of the weight curve. 
Additionally, genetic factors inherent to different avocado varieties may predispose them to exhibit either a single or double sigmoidal curve in weight development.
Given this and the availability of data from the literature, we propose to model dry matter via a sigmoidal growth model, and fruit weight via a double sigmoidal growth model.
Such models are defined in the next section.

The sampling design problem that we consider will be based on avocado growth in Australian conditions.
All previous research has been completed outside of Australia which suggests that the prior information, including the model, may be misspecified.
Thus, new methods are needed to derive Bayesian designs that are robust in such settings.

\section{Statistical modelling framework}

In this section, we describe the mathematical models used to describe fruit growth and development.
In addition, we describe the statistical modelling framework within which we draw inference.

\subsection{Ordinary differential equations}

Mechanistic models are mathematical models that are based on the underlying physical processes and mechanisms that govern the behaviour of a system.
These models attempt to capture the cause-and-effect relationships that underlie the behaviour of the system, rather than simply describing its observed behaviour.
ODE models are commonly used to describe mechanistic models because they are particularly well-suited for capturing the dynamics of physical systems that change over time \citep*{strogatz1994nonlinear}.

Throughout this paper, we consider ODEs to describe fruit growth, development and maturity as indicated by dry matter and weight.
The proposed mathematical functions for modelling growth include the Gompertz, Logistic, Richards and models based on the Weibull distribution, each with distinct parameterisations that impart specific characteristics of growth, forming families of curves \citep*{ratkowsky1983model}. These models have been commonly used to model the growth of fruits such as papaya fruit \citep*{salinas2019fruit,meza2011evaluacion}, peach fruit \citep{lescourret2011qualitree, miras2011qualitree, miras2013combined, miras2013assessment, bevacqua2019coupling}, pepper fruit \citep*{wubs2012model}, kiwi fruit \citep*{lescourret1998pollination}, summer squash \citep*{rodriguez2015growth}, tomato \citep{grange1993growth,adams2001effect}, cacao fruit \citep*{muniz2017nonlinear} and coffee berries \citep*{fernandes2017double}.
Given this and the work with avocado fruit in New Zealand \citep*{clark2003dry,dixon2006patterns} and South Africa \citep{moore1995effect}, we propose to consider such models as a basis to determine sampling designs for avocados in Australia.

An ODE model has the form,
$\mbox{d}y / \mbox{d}t = f(y,t)$,
where $y$ is the response at time $t$ and the solution of the ODE is the mean of the response.
A general sigmoidal growth model has the form,
$\mbox{d}y / \mbox{d}t = ryf(y)$,
where $y \geq0$ is the response at time $t \geq0$, $r>0$ is the growth rate, and $f(y)$ is a crowding function that accounts for the effects of crowding and competition, which diminish the net growth rate as $y$ increases \citep{jin2016stochastic}.
There are numerous options for the crowding function and decreasing functions are common choices for $f(y)$, $\mbox{d}f/\mbox{d}y<0$, with $f(\lambda)=0$, where $\lambda>0$ is the carrying capacity. 
Typical choices of $f(y)$ give different growth models, i.e., $f(y)=\log(\lambda/y)$ is the Gompertz model, $f(y)=1-(y/\lambda)$ is the Logistic model and $f(y)=1-(y/\lambda)^a$ is the Richards model, for some constant $a>0$.

\subsubsection{Ordinary differential equation to describe dry matter}
To describe dry matter, we consider two mathematical models of population growth with mean $E[{y}|{t},r,\lambda,y_0]$, which is a solution of an ODE.
They are,

Gompertz model:
\begin{equation} \label{G_Ey1}
\begin{split}
& \frac{\mbox{d}y}{\mbox{d}t} = r y \log \left(\frac{\lambda}{y} \right), \mbox{with solution} \\
& \mbox{E}[y|t,r,\lambda,y_0] = \lambda \exp \left( {-\log \left(\frac{\lambda}{y_0} \right) \exp{{{(-r{{t}})}}}} \right),
\end{split}
\end{equation}
Logistic model:
\begin{equation} \label{L_Ey1}
\begin{split}
& \frac{\mbox{d}y}{\mbox{d}t} = r y \left(1 - \frac{y}{\lambda} \right),  \mbox{with solution} \\
& \mbox{E}[y|t,r,\lambda,y_0] = { {\frac {\lambda y_0}{y_0 + {(\lambda - y_0}) \exp {{{(-r{t)}}}}}} },
\end{split}
\end{equation}

where $y \geq0$ is the response at time $t \geq0$. 
The growth rate is represented by $r>0$ and the carrying capacity is denoted by $\lambda>0$.
The initial response is $y_0$.
Then the responses can be modelled by, $y_i \sim \mbox{N} (E[y|t_i,r,\lambda,y_0],\sigma_e^2)$ for $i=1,2,\dots,n$ where $n$ is the number of data points and $\sigma_e^2$ is the residual variance.

\subsubsection{Ordinary differential equation to describe weight}
To describe weight, we consider the double Gompertz model to describe the two phases of growth. The model can be defined as follows,

Double Gompertz model:
\begin{equation} \label{G_Ey2}
\begin{split}
& \frac{\mbox{d}y_1}{\mbox{d}t} = ry_1 \log \left(\frac{\lambda_1}{y_1} \right), \mbox{with solution} \\
& \mbox{E}[y_1|t,r,\lambda_1,y_0] = \lambda_1 \exp \left({-\log \left(\frac{\lambda_1}{y_0} \right) \exp{{(-r{t)}}} }\right) ; 0<t \leq \eta ,\\
& \frac{\mbox{d}y_2}{\mbox{d}t} = r y_2   \log \left(\frac{\lambda_2}{y_2} \right), \mbox{with solution} \\
& \mbox{E}[y_2|t,r,\lambda_2,\mbox{E}[y_1|\eta,r,\lambda_1,y_0]] = \lambda_2 \exp \left ( {-\log \left(\frac{\lambda_2}{\mbox{E}[y_1|\eta,r,\lambda_1,y_0]} \right) \exp {{(-r({t-\eta))}}} } \right); 
\\ & \eta \leq t \leq T , 
\end{split}
\end{equation}

where $y_1 \geq0$ and $y_2 \geq0$ are the responses in the first and second phase, respectively. 
The carrying capacities of first and second phase are $\lambda_1>0$ and $\lambda_2>\lambda_1$, respectively. 
$\eta$ is the switch time from the first phase to the second phase and $T$ is the duration of the fruit growth.
Then the responses can be modelled by, $y_{1i} \sim \mbox{N} (E[y_1|t_{i},r,\lambda_1,y_0],\sigma_e^2)$ for $i=1,2,\dots,n_1$ and $y_{2i} \sim \mbox{N} (E[y_2|t_{i},r,\lambda_2,{\mbox{E}[y_1|\eta,r,\lambda_1,y_0]}],\sigma_e^2)$ for $i=1,2,\dots,n_2$ where $n_1$ and $n_2$ are number of data points in the first and second phase of growth, respectively.

\subsection{Flexible ordinary differential equations}

There are several approaches to allow a model to be more flexible, depending on the type of model and the data.
Among them, splines can be used to obtain a flexible representation of the solution of an ODE as they provide a smooth function that can approximate the true solution of the ODE.
Incorporating spline terms into ODE models provides a flexible and versatile approach to capture complex dynamics, particularly in cases where the underlying system is non-linear, has smooth transitions between states, or exhibits piecewise behaviour. 
However, as with any modelling approach, careful consideration should be given to the choice of spline basis, the number of knots, and other modelling parameters to ensure that the resulting model accurately represents the underlying system while avoiding overfitting.
In this section, we derive flexible ODEs through the use of a spline term with a polynomial basis function.

\subsubsection{Flexible model for dry matter}

In order to describe dry matter, we consider Gompertz and Logistic models with a spline model incorporated to allow flexibility in describing how dry matter evolves over time.
Specifically, these models can be defined as follows,

Flexible Gompertz model:
\begin{equation} \label{GS_Ey1}
\begin{split}
& \frac{\mbox{d}y}{\mbox{d}t} = r y \log \left( \frac{\lambda}{y} \right)  B'(t), \mbox{with solution}\\
& \mbox{E}[y|t,r,\lambda,y_0,\beta_0,\beta_1,\boldsymbol{\beta_{2}}] = \lambda \exp \left ( {-\log \left( \frac{\lambda}{y_0} \right) \exp {{(-rB(t))}}  }\right),
\end{split}
\end{equation}

Flexible Logistic model:
\begin{equation} \label{LS_Ey1}
\begin{split}
& \frac{\mbox{d}y}{\mbox{d}t} = r y  \left(1 - \frac{y}{\lambda} \right) B'(t), \mbox{with solution}\\
& \mbox{E}[y|t,r,\lambda,y_0,\beta_0,\beta_1,\boldsymbol{\beta_{2}}] =  \frac {\lambda y_0}{y_0 +{(\lambda - y_0}) \exp {{(-r{B(t))}}}},
\end{split}
\end{equation}
\sloppy where $B'(t) =  \left(\beta_0 + \beta_1 t + \sum_{k=1}^{K} {\beta_{2k}} \left( {t}^{k}-{\tau_k}\right) \right)$ is the derivative of $B(t) = \left(\beta_0 t + \beta_1 (\frac{t^2}{2}) + \sum\limits_{k=1}^{K} {\beta_{2k}} \left( \frac{{t}^{k+1}}{k+1}-{\tau_k}{t}\right) \right)$, $K$ is the number of knots, $\boldsymbol{\beta_{2}}=(\beta_{21},\beta_{22},\dots,\beta_{2K})$ is the vector of random effects, $\boldsymbol{\tau}=(\tau_1,\tau_2,\dots,\tau_K)$ is the vector of knots and $({t}^{k}-{\tau_k})$ is the spline basis function for $k=1,2,\dots,K$.

\subsubsection{Flexible model for weight}
To describe weight, we consider the double Gompertz model to describe the two phases of
growth with a spline term incorporated in both phases.
The model can be defined as follows,

Flexible double Gompertz model:
\begin{equation} \label{GS_Ey2}
\begin{split}
& \frac{\mbox{d}y_1}{\mbox{d}t} = r y_1  \log \left(\frac{\lambda_1}{y_1} \right) {B_{1}'(t)}, \mbox{with solution} \\
& \mbox{E}[y_1|t,r,\lambda_1,y_0,\beta_{01},\beta_{11},\boldsymbol{\beta_{3}}] = \lambda_1 \exp \left ({-\log \left(\frac{\lambda_1}{y_0} \right) \exp {{(-r{B_{1}(t))}} }} \right); 0<t \leq \eta ,\\
& \frac{\mbox{d}y_2}{\mbox{d}t} = ry_2 \log \left(\frac{\lambda_2}{y_2} \right) {B_{2}'(t)}, \mbox{with solution} \\
& \mbox{E}[y_2|t,r,\lambda_2,{\mbox{E}[y_1|\eta,r,\lambda_1,y_0,\beta_{01},\beta_{11},\boldsymbol{\beta_{3}}]},\beta_{02},\beta_{12},\boldsymbol{\beta_{4}}] = \\ 
& \lambda_2 \exp \left ({-\log \left(\frac{\lambda_2}{\mbox{E}[y_1|\eta,r,\lambda_1,y_0,\beta_{01},\beta_{11},\boldsymbol{\beta_{3}}]} \right) \exp {{(-r{(B_{2}(t)-\eta))}} }} \right); \eta \leq t\leq T ,
\end{split}
\end{equation}

\sloppy where ${B_{1}'(t)} = \left(\beta_{01}  + \beta_{11}t + \sum_{k_1=1}^{K_1} {\beta_{3k_1}} \left( {t}^{k_1}-{\tau_{k_1}}\right) \right)$ is the derivative of ${B_{1}(t)} = \left(\beta_{01} t + \beta_{11} (\frac{t^2}{2}) + \sum\limits_{k_1=1}^{K_1} {\beta_{3k_1}} \left( \frac{{t}^{k_1+1}}{k_1+1}-{\tau_{k_1}}t\right) \right)$ and ${B_{2}'(t)} = \left(\beta_{02} + \beta_{12}t + \sum_{k_2=1}^{K_2} {\beta_{4k_2}} \left( {t}^{k_2}-{\tau_{k_2}}\right) \right)$ is the derivative of ${B_{2}(t)} = \left(\beta_{02} t + \beta_{12} (\frac{t^2}{2}) + \sum\limits_{k_2=1}^{K_2} {\beta_{4k_2}} \left( \frac{{t}^{k_2+1}}{k_2+1}-{\tau_{k_2}}t\right) \right)$. 
Number of knots in the first and second phase of the model are $K_1$ and $K_2$, respectively. 
$\boldsymbol{\beta_{3}}=(\beta_{31},\beta_{32},\dots,\beta_{3{K_1}})$ is the vector of random effects of the first phase, $\boldsymbol{\beta_{4}}=(\beta_{41},\beta_{42},\dots,\beta_{4{K_2}})$ is the vector of random effects of the second phase, $\boldsymbol{\tau_{1}}=(\tau_{1},\tau_{2},\dots,\tau_{K_1})$ is the vector of knots in the first phase and $\boldsymbol{\tau_{2}}=(\tau_{1},\tau_{2},\dots,\tau_{K_2})$ is the vector of knots in the second phase. 
As discussed in Section 2, fruit weight can be obtained via repeated measures over time.
Therefore, to account for variation between and within fruit, a random effect could be incorporated on, for example, the growth rate.
Further details about this will be given in the next section.

\subsection{Bayesian Inference for ODE models}
This study will employ a Bayesian framework, whereby all inferences are based on the posterior distribution. 
Let $p(\boldsymbol{\theta})$ denote the prior distribution of the parameters $\boldsymbol{\theta}$ and $p(\boldsymbol{y}|\boldsymbol{\theta},\boldsymbol{t})$ is the likelihood of observing $\boldsymbol{y}$ given $\boldsymbol{\theta}$ and $\boldsymbol{t}$, then the posterior distribution of $\boldsymbol{\theta}$ can be defined as,
\begin{equation} \label{post}
p(\boldsymbol{\theta}|\boldsymbol{y},\boldsymbol{t}) = \frac{p(\boldsymbol{y}|\boldsymbol{\theta},\boldsymbol{t}) p(\boldsymbol{\theta})} {p(\boldsymbol{y})},
\end{equation} 
where $p(\boldsymbol{y})$ is the normalising constant.
A variety of different algorithms have been proposed to efficiently approximate the posterior distribution, as the product of $p(\boldsymbol{y}|\boldsymbol{\theta},\boldsymbol{t})$ and $p(\boldsymbol{\theta})$ typically does not yield a known distribution.
A widely adopted method for this purpose is Markov chain Monte Carlo, see \citet*{metropolis1953equation}.

The distribution of interest for the growth models discussed in Section 3.1.1 is the posterior distribution of $\boldsymbol{\theta}$, which can be defined as follows,
\begin{equation} \label{posteriorG1}
p(\boldsymbol{\theta}|\boldsymbol{y},\boldsymbol{t}) \propto p(\boldsymbol{y}|r, \lambda,\boldsymbol{t}) p(\boldsymbol{\theta}),
\end{equation}
where $\boldsymbol{\theta} =(r, \lambda, \sigma_e^2)$, $p(\boldsymbol{y}|r, \lambda, \boldsymbol{t})$ is the likelihood of observing $\boldsymbol{y}$ at time $\boldsymbol{t}$ given $r$ and $\lambda$.
For the growth model described in Section 3.1.2, the posterior distribution of $\boldsymbol{\theta}$ can be expressed as in Equation (\ref{posteriorG1}) where $\boldsymbol{\theta} =(r, \lambda_1, \lambda_2, \sigma_e^2)$.

The distribution of interest for the flexible growth models discussed in Section 3.2.1 is the posterior distribution of $\boldsymbol{\theta}$ and $\boldsymbol{b}$, which can be defined as follows,
\begin{equation} \label{posterior1}
p(\boldsymbol{\theta},\boldsymbol{b}|\boldsymbol{y},\boldsymbol{t}) \propto p(\boldsymbol{y}|\boldsymbol{\theta},\boldsymbol{b},\boldsymbol{t})p(\boldsymbol{b}|\sigma_{b}^2) p(\boldsymbol{\theta}),
\end{equation} 
where $ \boldsymbol{\theta} =(\boldsymbol{\theta_{\kappa}},\boldsymbol{\theta_{\gamma}})$, $ \boldsymbol{\theta_{\kappa}} =(r, \lambda, \beta_0, \beta_1)$, $ \boldsymbol{\theta_{\gamma}} =(\sigma_e^2, \sigma_{b}^2)$, $\boldsymbol{b} = (\beta_{21},\beta_{22},...,\beta_{2K})$ and $\sigma_{b}^2$ is the variance of the random effects $\boldsymbol{b}$.

For weight, there is the additional consideration of accounting for variability in growth rates within and between fruit.  
For this, we assume there are G fruits, with each fruit indexed by $g = 1,2,\dots ,G$. 
Then define the growth rate for each fruit as:
\begin{equation} \label{g}
\log(\boldsymbol{r_g}) = \log(r) + {\boldsymbol{b_g}} ,
\end{equation}
where $\boldsymbol{r_g} = (r_1,r_2,\dots ,r_G)$ are the growth rates of different fruits and $\boldsymbol{b_g}$ is the vector of random effects which is assumed to follow Normal distribution with mean 0 and variance $\sigma_{b_g}^2$.
Then the posterior distribution of the flexible growth model described in Section 3.2.2 can be described similar to Equation (\ref{posterior1})
where $\boldsymbol{\theta_\kappa} =(r, \lambda_1, \lambda_2, \beta_{01}, \beta_{11}, \beta_{02}, \beta_{12})$, $\boldsymbol{\theta_\gamma} =( \sigma_e^2, \boldsymbol{\sigma_{b}^2})$, $\boldsymbol{b} = (\boldsymbol{b_1},\boldsymbol{b_2},\boldsymbol{b_g})$, $\boldsymbol{b_1} = (\beta_{31},\beta_{32},\dots,\beta_{3K_1})$, $\boldsymbol{b_2} = (\beta_{41},\beta_{42},\dots,\beta_{4K_2})$ and $\boldsymbol{\sigma_{b}^2} = (\sigma_{b_1}^2, \sigma_{b_2}^2, \sigma_{b_g}^2)$. 
The variances of random effects $\boldsymbol{b_1}$ and $\boldsymbol{b_2}$ are $\sigma_{b_1}^2$ and $\sigma_{b_2}^2$, respectively.

\section{Bayesian design}

This section describes Bayesian design for the growth models and the optimisation algorithm that we used to find designs.
For growth models, a design $\boldsymbol{d}$ typically relates to the time $\boldsymbol{t}$ at which the data are to be collected. 
The aim of Bayesian design is to define $\boldsymbol{d}$ in order to achieve an experimental goal. 
The experimental goal is typically specified via a utility function which is defined in the next section.

\subsection{Utility function}

A utility function quantifies how well the goal of data collection would be achieved if design $\boldsymbol{d}$ were used to observe data $\boldsymbol{y}$. 
As the observed data is unknown, the expected value of the utility, evaluated over the prior predictive distribution is used to guide sampling.  
Specifically, a design is found by maximising the following expected utility:
\begin{equation} \label{utility}
\begin{split}
U(\boldsymbol{d}) & = \mbox{E}_{y,\theta,b}[u(\boldsymbol{d},\boldsymbol{y},\boldsymbol{\theta},\boldsymbol{b})] \\
 & = \int_{\boldsymbol{Y} }\int_{\boldsymbol{\Theta}} \int_{\boldsymbol{B}} u(\boldsymbol{d},\boldsymbol{y},\boldsymbol{\theta},\boldsymbol{b}) p(\boldsymbol{y},\boldsymbol{\theta},\boldsymbol{b}|\boldsymbol{d}) \mbox{d}\boldsymbol{b} \mbox{d}\boldsymbol{\theta} \mbox{d}\boldsymbol{y} \\
 & = \int_{\boldsymbol{Y} }\int_{\boldsymbol{\Theta}} \int_{\boldsymbol{B}} u(\boldsymbol{d},\boldsymbol{y},\boldsymbol{\theta},\boldsymbol{b}) p(\boldsymbol{\boldsymbol{y}}|\boldsymbol{\theta},\boldsymbol{b},\boldsymbol{d}) p(\boldsymbol{\theta},\boldsymbol{b}|\boldsymbol{d}) \mbox{d}\boldsymbol{b} \mbox{d}\boldsymbol{\theta} \mbox{d}\boldsymbol{y}.
 \end{split}
\end{equation} 
Unfortunately, in most cases, Equation (\ref{utility}) is analytically intractable so numerical methods are needed to approximate it.
The most commonly used approach for this is Monte Carlo integration in which the expected utility is approximated as follows,
\begin{equation} \label{appro_utility}
\hat U(\boldsymbol{d}) \approx \frac{1}{L} \sum_{l=1}^L u(\boldsymbol{d},\boldsymbol{y_l},\boldsymbol{\theta_l},\boldsymbol{b_l}),
\end{equation} 
where $\boldsymbol{y_l}, \boldsymbol{\theta_l}, \boldsymbol{b_l} \sim p(\boldsymbol{y}, \boldsymbol{\theta}, \boldsymbol{b})$ for  $l=1,2, \dots , L$.  
Here the value $L$ can be chosen based on the variability of the expected utility and time taken to evaluate the approximation. 
That is, $L$ can be selected so that the variability of the expected utility is relatively small and the compute time is feasible.
Then, a Bayesian design can be found by determining the design $\boldsymbol{d}$ that maximises the above approximation to the expected utility.

The Kullback-Liebler divergence (KLD) \citep*{kullback1951information} is a popular utility function in Bayesian design. 
It quantifies the divergence between the prior and posterior distributions of the parameters with larger differences suggesting that more has been learned from the observed data.
The KLD utility is,
\begin{equation} \label{priorpost}
U(\boldsymbol{d},\boldsymbol{y}) = \int_{\boldsymbol{\Theta}} \int_{\boldsymbol{B}} \log \left( \frac{p(\boldsymbol{\theta},\boldsymbol{b}|\boldsymbol{y},\boldsymbol{d})}{p(\boldsymbol{\theta},\boldsymbol{b})} \right) p(\boldsymbol{\theta},\boldsymbol{b}|\boldsymbol{y},\boldsymbol{d}) \mbox{d}\boldsymbol{b} \mbox{d}\boldsymbol{\theta}.
\end{equation} 
Evaluating KLD for a design and data set thus requires a posterior distribution.  
From Equation (\ref{appro_utility}), it can be seen that this evaluation is repeated $L$ times to approximate the expected utility.  
This is also repeated many times in search for the optimal design (discussed below).  
This renders algorithms like MCMC impractical for use in realistically sized Bayesian design problems. 
Therefore, there is a need for an efficient method to compute the posterior distribution.
In this paper, we consider the Laplace approximation as such a method, and the next section goes into detail about this.

\subsection{An efficient approximation to the posterior distribution}

Computing the Monte Carlo approximation to the expected utility function, as shown in Equation (\ref{appro_utility}), requires approximating or sampling from a large number of posterior distributions.
That is, a large number of prior predictive samples are generated and the utility is evaluated for each data set. 
The average of these utilities is then taken as the approximation to the expected utility.
For computational efficiency, the Laplace approximation has been considered in Bayesian experimental design contexts \citep*{ryan2003estimating}.
The Laplace approximation yields a multivariate Normal approximation to the posterior based on locating the mode of the posterior distribution and evaluating the curvature of the posterior distribution around this mode.
Such an approximation to $p(\boldsymbol{\theta}|\boldsymbol{y},\boldsymbol{d})$, follows a multivariate Normal distribution with mean $\boldsymbol{\theta^*}$ and covariance $\boldsymbol{A(\theta^* )}^{-1}$
where,
\begin{equation} \label{theta}
\begin{split}
& \boldsymbol{\theta^*} = \arg \max_{\boldsymbol{\theta}} \{ \log p(\boldsymbol{y}|\boldsymbol{\theta},\boldsymbol{d}) + \log p(\boldsymbol{\theta}) \} \; \mbox{and} \\
& \boldsymbol{A(\theta^*)} = \left. \frac{-\partial^2\{ \log p(\boldsymbol{y}|\boldsymbol{\theta},\boldsymbol{d}) + \log p(\boldsymbol{\theta}) \}}{\partial \boldsymbol{\theta} \partial \boldsymbol{\theta'}} \right|_{\boldsymbol{\theta} = \boldsymbol{\theta^*}}.
\end{split}
\end{equation} 
The Laplace approximation requires evaluating the full data likelihood.
For the flexible growth models described above, this can be obtained by integrating out the random effects as follows,
\begin{equation} \label{fullLikelihood}
p(\boldsymbol{y}|\boldsymbol{\theta},\boldsymbol{d}) = \int_{\boldsymbol{B}} p(\boldsymbol{y}|\boldsymbol{\theta},\boldsymbol{b},\boldsymbol{d}) p(\boldsymbol{b}|\boldsymbol{\sigma_{b}^2}) \mbox{d}\boldsymbol{b}.
\end{equation} 

Once $\boldsymbol{\theta^*}$ is obtained, the posterior distribution of the random effect parameters $\boldsymbol{b}$ can also be efficiently approximated.
This leverages the conditional independence between the model parameters and random effects, allowing us to approximate the marginal posterior distribution of the random effect parameters.
Let the random variable representing the marginal posterior of $\boldsymbol{b}$ given $\boldsymbol{\theta^*}$ be denoted as $\boldsymbol{b_{\theta}^*}$, then, the marginal posterior $p(
\boldsymbol{b_{\theta^*}}|\boldsymbol{y},\boldsymbol{d})$, follows a multivariate Normal distribution with mean $\boldsymbol{b_{\theta^*}^*}$ and covariance $\boldsymbol{A(b_{\theta^*}^*})^{-1})$
where,
\begin{equation} \label{alpha}
\begin{split}
& \boldsymbol{b_{\theta^*}^*} = \arg \max_{\boldsymbol{{b_{\theta^*}}}} \{ \log p(\boldsymbol{y}|\boldsymbol{\theta^*},\boldsymbol{b_{\theta^*}},\boldsymbol{d}) + \log p(\boldsymbol{b_{\theta^*}}|\boldsymbol{{\sigma_{b}^{2}}^*}) \} \; \mbox{and} \\
& \boldsymbol{A(b_{\theta^*}^*)} = \left.  \frac{-\partial^2 \{ \log  p(\boldsymbol{y}|\boldsymbol{\theta^*},\boldsymbol{b_{\theta^*}},\boldsymbol{d}) + \log p(\boldsymbol{b_{\theta^*}}|{\boldsymbol{\sigma_{b}^{2}}^*}) \}}{\partial {\boldsymbol{b_{\theta^*}}} \partial \boldsymbol{b_{\theta^*}'}} \right|_{\boldsymbol{b_{\theta^*}} = \boldsymbol{b_{\theta^*}^*}}.
\end{split}
\end{equation} 

If the prior distribution follows a multivariate Normal distribution with the prior mean $\boldsymbol{\mu_0}$, posterior mean $\boldsymbol{\mu_1}$, prior covariance matrix $\boldsymbol{\Sigma_0}$ and posterior covariance matrix $\boldsymbol{\Sigma_1}$, then the KLD can be computed analytically using the following formula,
\begin{equation} \label{KLD}
U(\boldsymbol{d},\boldsymbol{y}) = 0.5 \left( \mbox{tr} (\boldsymbol{\Sigma_0}^{-1} \boldsymbol{\Sigma_1}) + (\boldsymbol{\mu_0} - \boldsymbol{\mu_1})^{T} \boldsymbol{\Sigma_0}^{-1} (\boldsymbol{\mu_0} - \boldsymbol{\mu_1}) -M - \ln \left(\frac{\det(\boldsymbol{\Sigma_1})}{\det(\boldsymbol{\Sigma_0})} \right) \right),
\end{equation} 
where $M$ is the number of parameters. 

\begin{algorithm}
\caption{Approximating the expected utility of a design}\label{alg:expUtility}
\SetAlgoLined
Initialise the prior information $p(\boldsymbol{\theta},\boldsymbol{b})$ and the design $\boldsymbol{d}$.\\
\For{$l = 1$ to $L$}
{
    Simulate $\boldsymbol{\theta_l}$ and $\boldsymbol{b_l}$ from prior $p(\boldsymbol{\theta},\boldsymbol{b})$. \\
    Simulate data $\boldsymbol{y_l}$ at the design $\boldsymbol{d}$ based on $\boldsymbol{\theta_l}$ and $\boldsymbol{b_l}$ from the assumed model outlined in Equations (\ref{GS_Ey1}), (\ref{LS_Ey1}), and (\ref{GS_Ey2}).\\
    Approximate $\boldsymbol{\theta_l^*}$ and $\boldsymbol{A(\theta_l^*)}$ for the simulated data using Equation (\ref{theta}). \\
    Approximate $(\boldsymbol{b_{\theta^*}^*})_l$ and the Hessian matrix $\boldsymbol{A((\boldsymbol{b_{\theta^*}^*})_l)}$ using Equation (\ref{alpha}) based on $\boldsymbol{\theta_l^*}$. \\
    Set the joint posterior, $p(\boldsymbol{\theta},\boldsymbol{b}|\boldsymbol{y},\boldsymbol{d}) \sim \mbox{MVN} ((\boldsymbol{\theta^*},(\boldsymbol{b_{\theta^*}^*})_l),(\boldsymbol{\Sigma_1})_l)$, where
    \[
    (\boldsymbol{\Sigma_1})_l =
    \begin{pmatrix}
        \boldsymbol{A(\theta_l^*)}^{-1} & 0 \\
        0 & \boldsymbol{A((\boldsymbol{b_{\theta^*}^*})_l)}^{-1}
    \end{pmatrix}
    \] \\
    Approximate the KLD utility $U_D(\boldsymbol{d},\boldsymbol{y_l})$ using Equation (\ref{KLD}). 
}
Approximate the expected utility 
$\hat{U}(\boldsymbol{d}) = \frac{1}{L} \sum\limits_{l=1}^L U_D(\boldsymbol{d},\boldsymbol{y_l})$.
\end{algorithm}

To summarise, our approach for approximating the expected utility is outlined in Algorithm \ref{alg:expUtility}.
The process is initialised by specifying the prior information and the design (line 1).
Next, model parameters are drawn (line 3), followed by generating a prior predictive sample (line 4).
Given these values, the posterior distribution of parameter $\boldsymbol{\theta_l}$ can be found by evaluating the full data likelihood using Equation (\ref{fullLikelihood}). 
However exact integration is typically not feasible, therefore Monte Carlo methods can be used to obtain an approximate solution, as outlined below,
\begin{equation} \label{mc}
p(\boldsymbol{y}|\boldsymbol{\theta},\boldsymbol{d}) \approx \frac{1}{E} \sum_{e=1}^{E} \prod_{i=1}^n p(\boldsymbol{y}|\boldsymbol{\theta},\boldsymbol{b_e},{d_i}),
\end{equation} 
where $E$ is sufficiently large.
Then the Laplace approximation can be employed to derive the posterior distribution of $\boldsymbol{\theta_l}$ (line 5) using the likelihood approximation in Equation (\ref{mc}). 
Then, incorporating an additional Laplace approximation enables the approximation of the posterior distribution of $\boldsymbol{b_{\theta^*}}$ (line 6).
The joint posterior distribution of $\boldsymbol{\theta}$ and $\boldsymbol{b}$ can then be estimated (line 7) from the output of lines 5 and 6, thereby allowing for the evaluation of the utility function (line 8).
After this has been repeated $L$ times, the expectation is approximated.

\subsection{Optimisation algorithm}

After obtaining an approximation of the expected utility, our next step is to find the design that maximises this approximation.
The coordinate-exchange algorithm (CE) proposed by \citet*{meyer1995coordinate} can be used to explore a discrete set of design points.
This algorithm starts with a random design and then optimises one design point at a time.
When optimising a design point, the given design point is substituted with an alternative design point and the expected utility is evaluated.
If this `new' design gives a larger expected utility value, then this design point is retained in the design. 
Otherwise, the original design will remain unchanged. 
This procedure is repeated for each alternative design point and continues until the design has been cycled through multiple times and/or there is no improvement in the expected utility.

Unfortunately, the CE algorithm can be expensive to deploy in settings where there is a continuous design space.
As an alternative, \citet*{overstall2017bayesian} introduced an approximate coordinate exchange (ACE) algorithm where a Gaussian process emulator is optimised for each one-dimensional optimisation within the CE algorithm, based on a relatively small number of expected utility values. 
The ACE algorithm has been progressively utilised in the literature within a variety of different applications \citep*{woods2017bayesian,overstall2018approach,dehideniya2018optimal,overstall2020bayesian,overstall2019bayesian}.
Accordingly, we adopt this optimisation algorithm to search for robust designs in our motivating application.

\section{Applications}
Our proposed methodology for finding robust Bayesian designs is applied below to determine sampling designs for two indicators of avocado growth and development.
These are fruit dry matter and fruit weight.
For each indicator, we find sampling designs as outlined in Section 4, and evaluate the robustness properties of these designs by considering potential alternative prior information to describe fruit growth.
Here, we largely consider misspecification in regards to the data generating model.
To evaluate such robustness, designs were also found under the Richards and Weibull growth models as alternatives to Gompertz and Logistic growth models, and the efficiency of our robust designs was evaluated with respect to these alternative models.
The efficiency of each design was evaluated as follows,
\begin{equation} \label{eff}
\begin{split}
E(\boldsymbol{d}) = \frac{U_{dgm}(\boldsymbol{d})}{U_{dgm}(\boldsymbol{d^*})} ,\\
\end{split}
\end{equation} 
where $U_{dgm}$ represents the expected utility under an assumed data generating model and $\boldsymbol{d^*}$ is the optimal design under this data generating model. 
Then, $E(\boldsymbol{d})$ is the amount of information that is expected to be obtained from $\boldsymbol{d}$ compared to $\boldsymbol{d^*}$.

To find our robust designs, we ran the ACE algorithm (Section 3.4.3) from 10 different randomly generated starting designs.
Then the expected utility was then re-evaluated for each of these 10 designs with a large value of $L$, and the design with the largest of these values was selected as optimal.

\subsection{Robust sampling designs for dry matter}
To derive robust Bayesian designs for dry matter, let $D$ be the design space of $\boldsymbol{d}$. 
We consider $D \in [0,30]^n$ for this example which denotes a 30 week period relating to the rapid growth phase after full bloom 
\citep*{valmayor1967cellular,lee1983maturity}.
As a basis for forming a robust design, consider the following flexible growth model:
\begin{equation} \label{model1}\boldsymbol{y}|r,\lambda,\beta_0,\beta_1,\boldsymbol{b} \sim  \mbox{N}(\mbox{E}[\boldsymbol{y}|\boldsymbol{t},r,\lambda,\beta_0,\beta_1,\boldsymbol{b},y_0],\sigma^2_e) \\
\end{equation} 
where $\boldsymbol{b} = (\beta_{21},\beta_{22},\dots,\beta_{2K})$ based on a given value for $K$.

The following priors were considered.
In terms of the mean values, these were based on the results of \citet*{clark2003dry} from avocados collected in New Zealand and reasonable values that might be anticipated to be observed.  
For example, typically dry matter for avocados is within the range of 0 to 0.3, so the prior mean for the log carrying capacity was set to log 0.3.
Similarly, the prior variances were selected to provide reasonable coverage of plausible parameter values.
\begin{equation} \label{prior1}
\begin{split}
& \log(r) \sim \mbox{N}(\log(0.1),0.1^2) \\
& \log(\lambda) \sim \mbox{N}(\log(0.3),0.1^2) \\
& \beta_0 \sim \mbox{N}(0.1,0.1^2) \\
& \beta_1 \sim \mbox{N}(0.1,0.1^2) \\
& \log(\sigma_e) \sim \mbox{N}(\log(1),0.05^2) \\
& \log(\sigma_{b}) \sim \mbox{N}(\log(\overline{\sigma_{b}}) ,0.4^2) \\
& \boldsymbol{b} \sim \mbox{N}(0,\overline{\sigma_b}^2).
\end{split}
\end{equation} 
Here, $\overline{\sigma_{b}}$ relates to the variance of the random effects within the spline term.  Throughout, we find designs based on a range of values of $\overline{\sigma_{b}}$ and $K$ (see Table~\ref{form1}) such that we can explore properties of the designs under different levels of model flexibility.

\begin{figure}[ht!]
    \centering
    \includegraphics[scale=0.43]{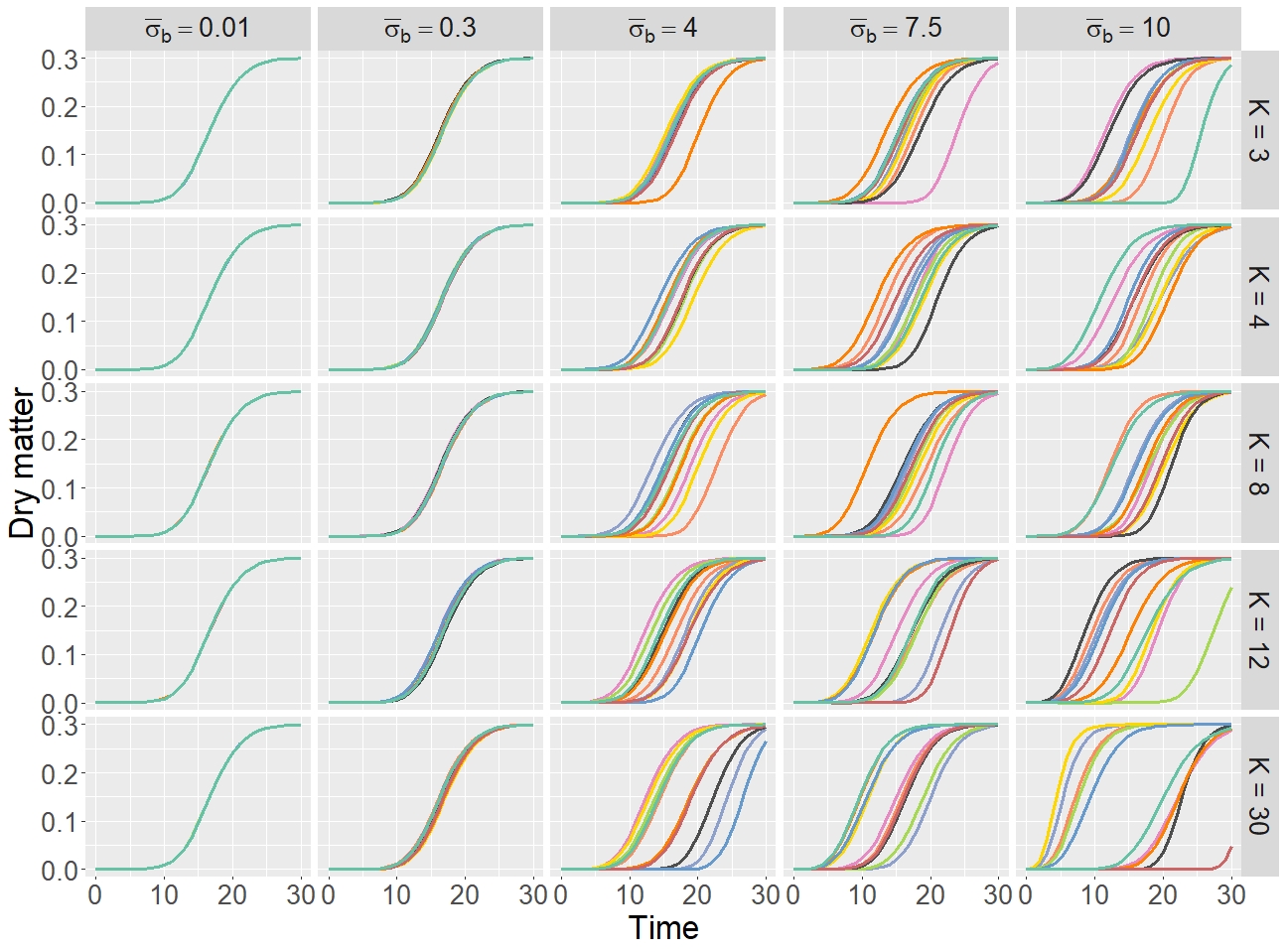}
    \caption{Ten potential realisations that could be captured by a Gompertz model with fixed values of $r$, $\lambda$, $\beta_0$ and $\beta_1$ where each color represents a different realisation.}
    \label{Realisations1}
\end{figure}

To illustrate the range of potential relationships between $\boldsymbol{t}$ and $\mbox{E}[\boldsymbol{y}|\boldsymbol{t},r,\lambda,\beta_0,\beta_1,\boldsymbol{b},y_0]$ that could be observed
under the model described in Equation (\ref{model1}), Figure \ref{Realisations1} presents realisations for different values of $\overline{\sigma_{b}}$ and $K$, assuming a Gompertz growth model for the expected value and fixed values of $r$, $\lambda$, $\beta_0$ and $\beta_1$.
From Figure \ref{Realisations1}, when $\overline{\sigma_{b}}$ is relatively low, only very small deviations were observed for all $K$. 
The flexibility of the model increases with $\overline{\sigma_{b}}$ where moderate flexibility appears when $\overline{\sigma_{b}} = 7.5$ and extreme flexibility appears when $\overline{\sigma_{b}} = 10$.
The flexibility of the model also seems to be associated with $K$, showing less flexibility when $K=4$ compared to larger values.

\begin{table}[h!]
  \centering
    \caption{Four levels of flexibility defined by different values for $\overline{\sigma_{b}}$ and $K$ for model for dry matter.} \label{form1}
\begin{tabular}{ |p{4cm}|p{1.5cm}|p{1.5cm}|  }
 \hline
Model formulation & $\overline{\sigma_{b}}$ &  $K$ \\ 
 \hline
       Very low flexibility & 0.01  & 3 \\
      Low flexibility   & 0.3 & 4 \\
      Medium flexibility  & 7.5 & 12 \\
      High flexibility  & 10.0 & 30 \\
 \hline
\end{tabular}
\end{table}

\begin{figure}[ht!]
    \centering
    \includegraphics[scale=0.38]{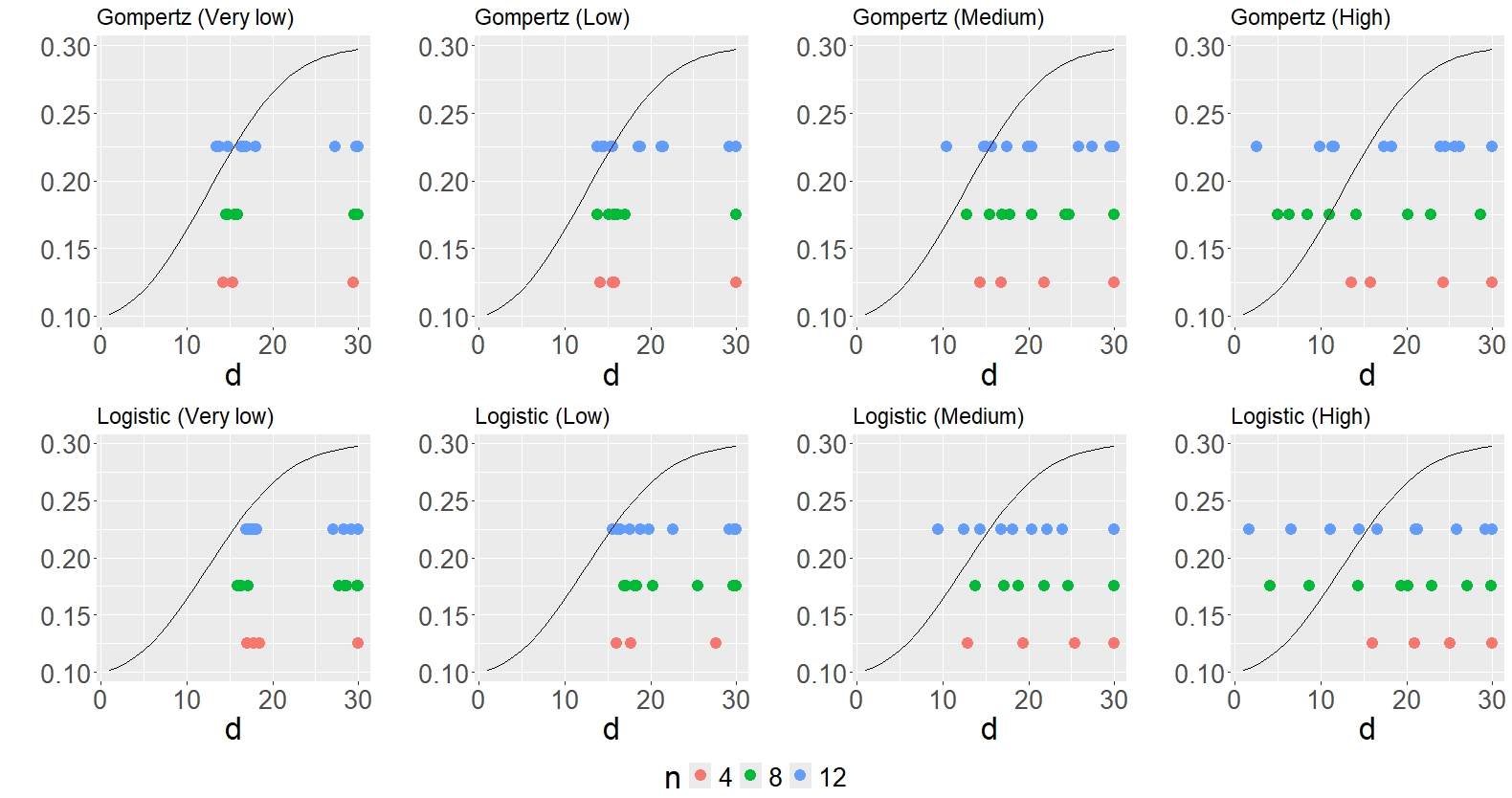}
    \caption{Optimal design points obtained for the Gompertz model (top) and Logistic model (bottom) under different model formulations.}
    \label{opt1}
\end{figure}

To explore the properties of our proposed approach, robust designs were found for the Gompertz and Logistic growth models under each model configuration given in Table~\ref{form1} and for $n$ = 4, 8 and 12.
The resulting designs are shown in Figure \ref{opt1}.
When the flexibility is very low, the design points appear to be more clustered around areas of the design space that would be useful for estimating the growth rate (i.e., $r$) and the carrying capacity (i.e., $\lambda$).
As the flexibility increases, the design points become more spread out, presumably because this allows the design to capture a wider range of dynamics.
This is particularly noticeable under the highly flexible model and $n$=12, as the design points essentially cover the entire design space.

\begin{figure}[htbp!]
    \centering
    \includegraphics[scale=0.34]{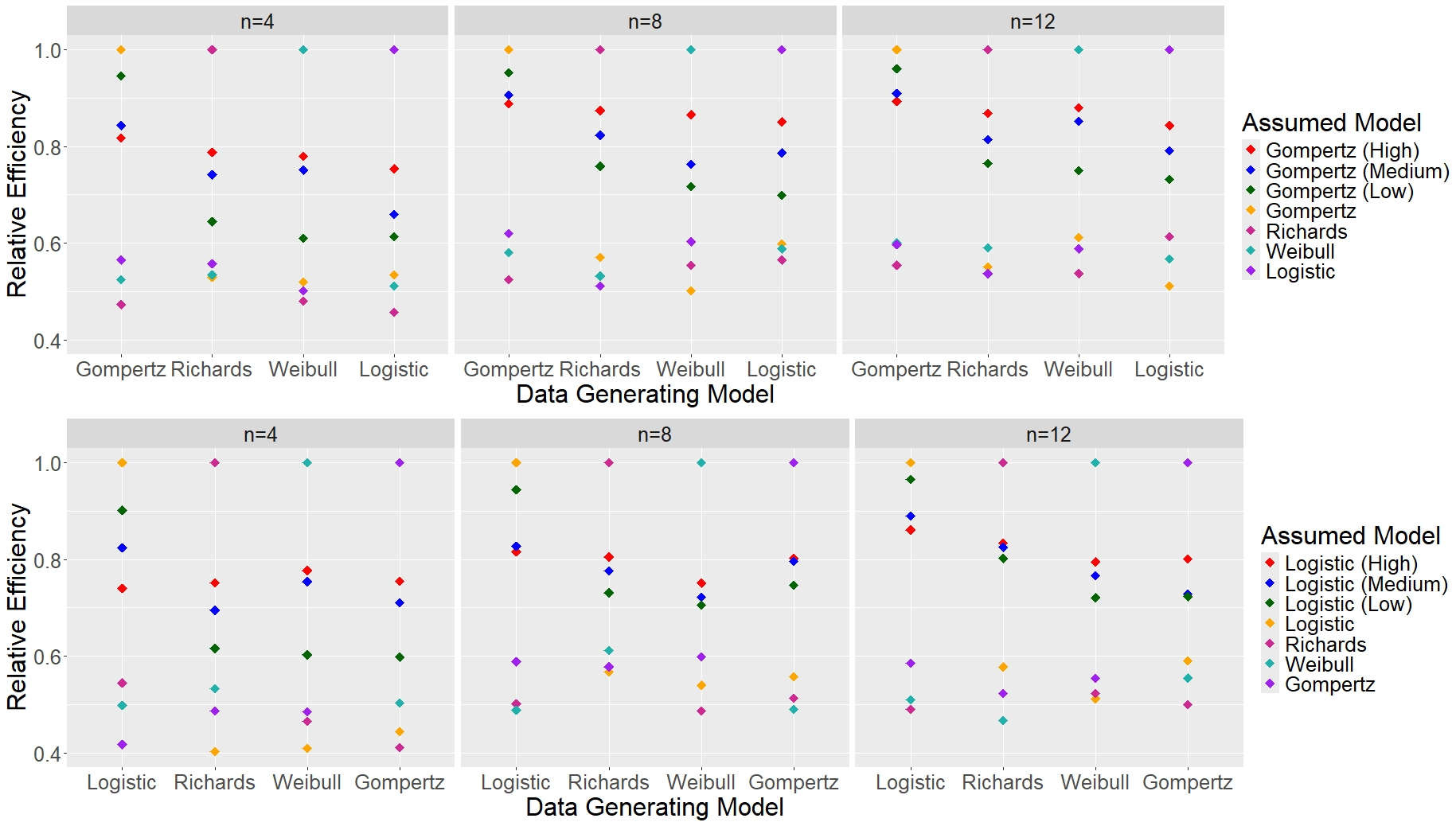}
    \caption{Relative efficiencies $E(\boldsymbol{d})$ of the optimal designs based on an assumed model compared to optimal designs based on the data generating model.}
    \label{re1}
\end{figure}

Figure \ref{re1} shows the efficiencies of the designs from Figure \ref{opt1} with respect to alternative ODE models for $n$ = 4, 8 and 12. 
$E(\boldsymbol{d}) = 1$ when the relevant data generating model is correctly assumed.
However, if this assumption is incorrect, the loss in efficiency can be substantial.
From Figure \ref{re1}, it is evident that the designs based on both the Gompertz model and Logistic model with high flexibility remain highly efficient under the alternative models.
That is, given a model with very low flexibility, the robust designs are around 40\% to 60\% efficient under alternative models whereas for a model with high flexibility, the robust designs are more than 70\% efficient when $n$ = 4 and more than 80\% efficient when $n$ = 8 and 12 under alternative models. 
This suggests that, within the setting of this application, the efficiency of the designs increases with the flexibility of the model.

\subsection{Robust sampling designs for fruit weight}

Next, we consider deriving robust Bayesian designs for fruit weight. 
We consider $D \in [0,350]^n$ which denotes a 350 day period relating to avocados being sampled up until the typical harvest period for producers \citep*{yahia2011avocado}.
As a basis for forming a robust design, consider the following flexible growth model:
\begin{equation} \label{model2}
\begin{split}
& \boldsymbol{y}|r,\lambda_1,\lambda_2, \beta_{01},\beta_{11},\beta_{02},\beta_{12}, \boldsymbol{b_1}, \boldsymbol{b_2}, \boldsymbol{b_g}, \eta \sim \\
& \mbox{N}(\mbox{E}[\boldsymbol{y}|\boldsymbol{t},r,\lambda_1,\lambda_2,\beta_{01},\beta_{11},\beta_{02},\beta_{12}, \boldsymbol{b_1}, \boldsymbol{b_2}, \boldsymbol{b_g}, \eta,y_0],\sigma^2_e),
\end{split}
\end{equation}

where $\boldsymbol{b_1} = (\beta_{31},\beta_{32},\dots,\beta_{3K_1})$, $\boldsymbol{b_2} = (\beta_{41},\beta_{42},\dots,\beta_{4K_2})$ and $\boldsymbol{b_g} = (\beta_{r1},\beta_{r2},\dots,\beta_{rG})$ based on given values for $K_1$, $K_2$ and $G$.

Based on the above model, the following priors were adopted.  
In terms of the prior means, these were determined based on the results of \citet*{dixon2006patterns} for studies conducted in New Zealand and \citet*{moore1995effect} in South Africa and through an assessment of reasonable values that might be anticipated to be observed.  
Further, the variances were determined such that the prior covered plausible ranges of parameter values.
\begin{equation} \label{prior2}
\begin{split}
& \log(r) \sim \mbox{N}(\log(0.02),0.1^2) \\
& \log(\lambda_1) \sim \mbox{N}(\log(150),0.1^2) \\
& \log(\lambda_2) \sim \mbox{N}(\log(200),0.1^2) \\
& \log(t_s) \sim \mbox{N}(\log(200),0.1^2) \\
& \beta_{01} \sim \mbox{N}(1,0.1^2) \\
& \beta_{11} \sim \mbox{N}(0.01,0.1^2) \\
& \beta_{02} \sim \mbox{N}(1,0.1^2) \\
& \beta_{12} \sim \mbox{N}(0.01,0.1^2) \\
& \log(\sigma_e) \sim \mbox{N}(\log(20),0.1^2) \\
& \log(\sigma_{b_1}) \sim \mbox{N}(\log(\overline{\sigma_{b_1}}) ,0.1^2) \\
& \log(\sigma_{b_2}) \sim \mbox{N}(\log(\overline{\sigma_{b_2}}) ,0.1^2) \\
& \log(\sigma_{b_g}) \sim \mbox{N}(\log(0.2) ,0.1^2) \\
& \boldsymbol{b_1} \sim \mbox{N}(0,{\overline{\sigma_{b_1}}}^2) \\
& \boldsymbol{b_2} \sim \mbox{N}(0,{\overline{\sigma_{b_2}}}^2) \\
& \boldsymbol{b_g} \sim \mbox{N}(0,0.2^2)
\end{split}
\end{equation} 
where $\overline{\sigma_{b_1}}$, $\overline{\sigma_{b_2}}$, $K_1$ and $K_2$ were used to specify a model with different levels of flexibility.
We assumed four different model formulations based on the Gompertz model and Logistic model with different $\overline{\sigma_{b_1}}$, $\overline{\sigma_{b_2}}$, $K_1$ and $K_2$ values, see Table~\ref{form2}. 

\begin{figure}[ht!]
    \centering
    \includegraphics[scale=0.4]{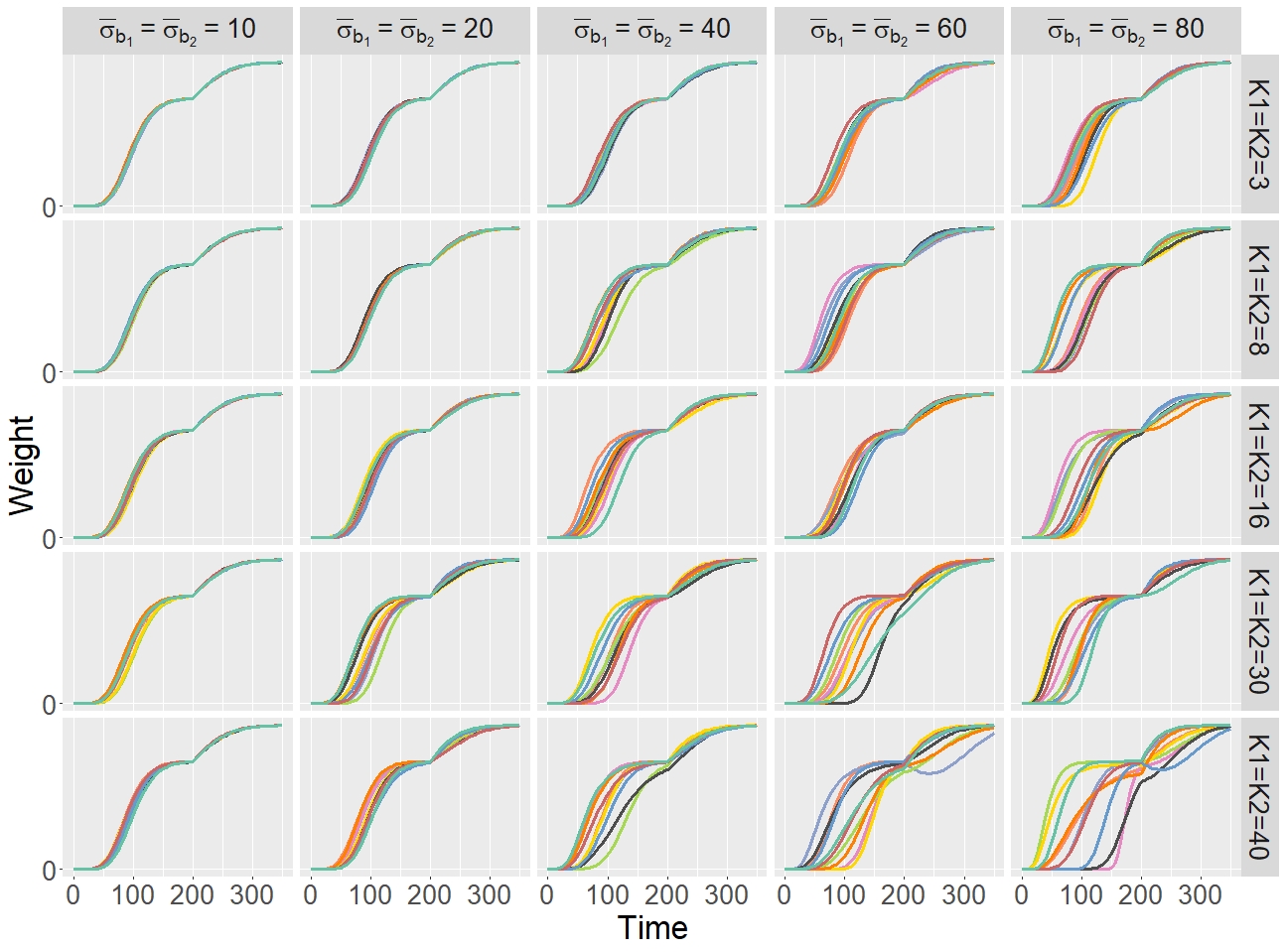}
    \caption{Ten potential realisations that could be captured by Gompertz model with fixed values for $r$, $\lambda_1$, $\lambda_2$, $t_s$, $\beta_{01}$, $\beta_{11}$, $\beta_{02}$ and $\beta_{12}$ where each color represents a different realisation.}
    \label{Realisations2}
\end{figure}

To provide insight into the flexibility that could be observed based on the above model and parameter values given in Table~\ref{form2}, potential realisations of $\mbox{E}[\boldsymbol{y}|\boldsymbol{t},r,\lambda_1,\lambda_2,\beta_{01},\beta_{11},\beta_{02},\beta_{12}, \boldsymbol{b_1}, \boldsymbol{b_2}, \boldsymbol{b_g}, \eta,y_0]$ are shown in Figure \ref{Realisations2} with fixed values for $r$, $\lambda_1$, $\lambda_2$, $t_s$, $\beta_{01}$, $\beta_{11}$, $\beta_{02}$ and $\beta_{12}$.
When $\overline{\sigma_{b_1}}$ and $\overline{\sigma_{b_2}}$ are relatively low, only very small deviations were observed for all $K_1$ and $K_2$. 
The flexibility of the model increases with $\overline{\sigma_{b_1}}$ and $\overline{\sigma_{b_2}}$ where moderate flexibility appears when $\overline{\sigma_{b_1}} = \overline{\sigma_{b_2}} = 40$ and extreme flexibility appears when $\overline{\sigma_{b_1}} = \overline{\sigma_{b_2}} = 80$ .
The flexibility of the model seems to be associated with $K_1$ and $K_2$, showing less flexibility when $K_1 = K_2 =8$ compared to larger values.

\begin{table}[h!]
  \centering
    \caption{Four different model formulations for model for fruit weight.} \label{form2}
\begin{tabular}{| p{4cm} | p{1cm} | p{1cm} | p{1cm} | p{1cm} | } 
 \hline
 Model formulation & $\overline{\sigma_{b_1}}$ &  $\overline{\sigma_{b_2}}$ & $K_1$ & $K_2$ \\   
 \hline
      Very low flexibility & 10 & 10 & 3 & 3 \\
      Low flexibility   & 20 & 20 & 16 & 16 \\
      Medium flexibility  & 60 & 60 & 30 & 30 \\
      High flexibility  & 80 & 80 & 40 & 40 \\
 \hline
\end{tabular}
\end{table}

\begin{figure}[ht!]
    \centering
    \includegraphics[scale=0.4]{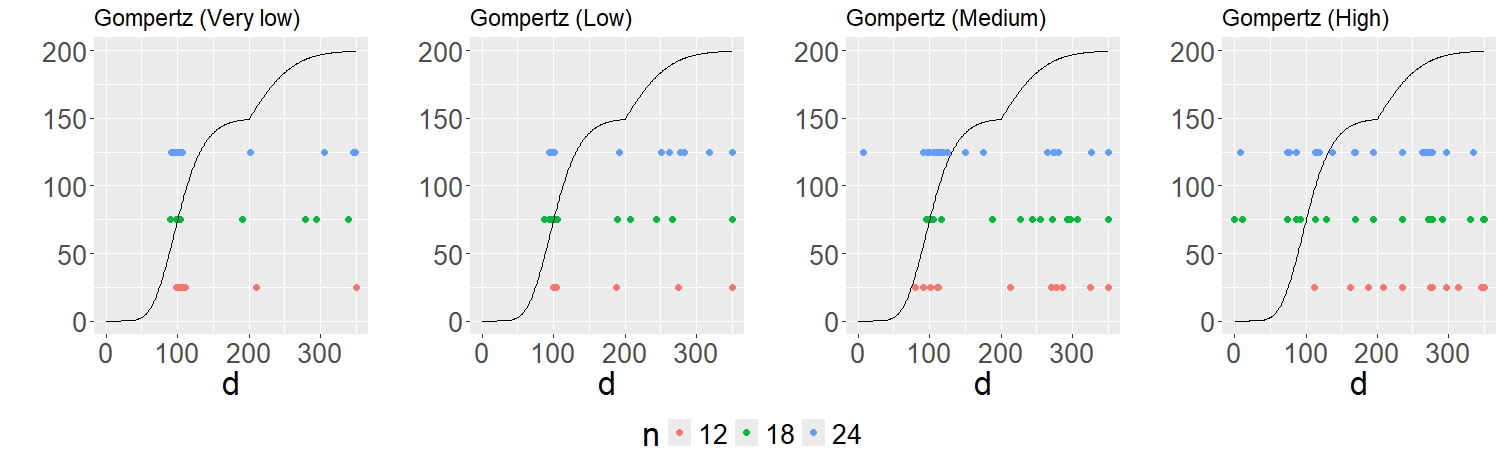}
    \caption{Optimal design points obtained for the Gompertz model under different model formulations.}
    \label{opt2}
\end{figure}

To explore designs under the different flexible models, we found an optimal design for each parameter configuration as given in Table~\ref{form2} with $n$ = 12, 18 and 24.
The resulting designs are shown in Figure \ref{opt2}, which shows that the design points appear to be more clustered around the areas of the design space that would be useful for estimating the parameters $r$, $\lambda_1$, $\lambda_2$ and $\eta$ when the model has very low flexibility.
As the flexibility increases, the design points become more spread out, presumably because this allows the design to efficiently capture a wider range of potential growth rates and carrying capacities.

\begin{figure}[htbp!]
    \centering
    \includegraphics[scale=0.34]{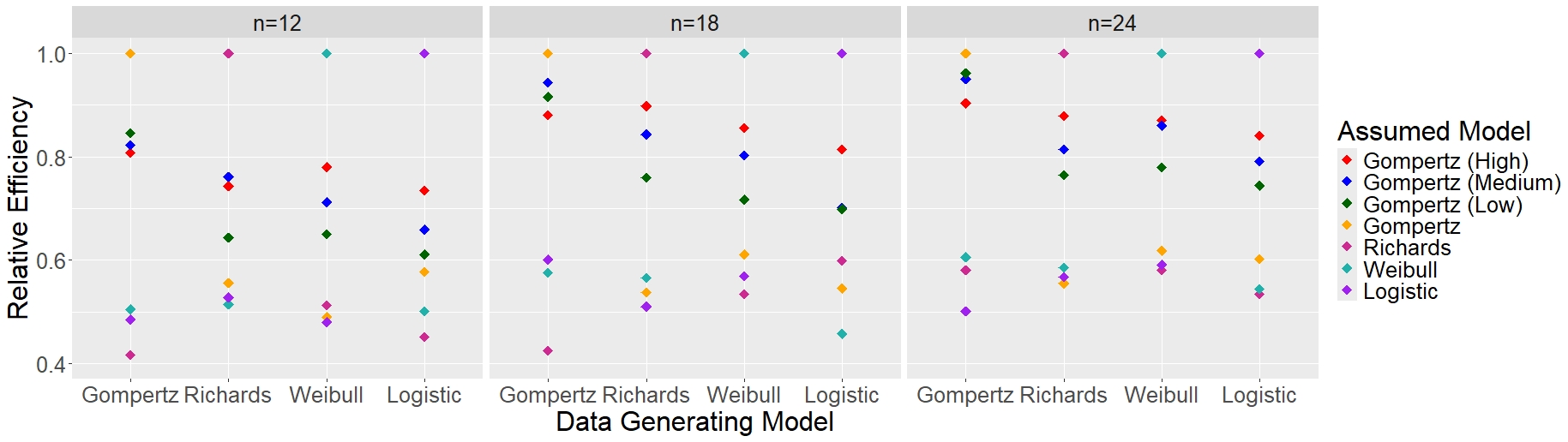}
    \caption{Relative efficiencies $E(\boldsymbol{d})$ of the optimal designs based on the Gompertz model in relation to the optimal designs based on an assumed model compared to optimal designs based on the data generating model.}
    \label{re2}
\end{figure}

Figure \ref{re2} shows the efficiencies for the designs shown in Figure \ref{opt2} under alternative growth models when $n$ = 12, 18 and 24.
The designs found based on the flexible double Gompertz model under the model with high flexibility remain highly efficient under the alternative models.
It can be seen that, under very low flexibility, the robust designs are around 40\% to 60\% efficient and under very high flexibility, the robust designs are more than 70\% efficient when $n$ = 12 and more than 80\% efficient when $n$ = 18 and 24 under alternative models. 
This suggests that, as with the designs for dry matter, the efficiency of the designs increases with the flexibility of the model.

\section{Discussion}

In this paper, we propose a Bayesian design approach that aims to find designs that are robust to the misspecification of prior information.
The main novelty is that designs are found under a class of models with associated prior knowledge, rather than being found under a single model with a prior on the parameter values.
The class of models considered here was formed through including a spline term into an ODE, and this was exploited to provide robust designs.
This work was motivated by a real-world design problem in agriculture where we demonstrated the robustness properties of our designs.
Such a property was highly desirable in this setting as it seemed unlikely that we could correctly specify the prior information based on studies conducted overseas.

Here our goal was to derive designs to improve avocado growth and show how flexible Gompertz and Logistic models can be used to form robust designs.
In the application, we explored the effect the model flexibility had on the subsequent form of the design by considering four different model formulations; very low flexibility, low flexibility, medium flexibility and high flexibility.
As shown, these flexible models remained efficient for parameter estimation under alternative models, and this robustness appeared to increase, in general, as the flexibility of the model increased.
Overall, it seemed that our proposed approach was able to provide efficient designs under models that were not specifically considered \textit{a priori}, and thus we suggest this is useful, in general, in Bayesian design.

One of the limitations of this work is that we assume that the data is normally distributed, and again this may not be reasonable in practice. 
To overcome this, one could consider a general Bayesian framework \citep{bissiri2016general} which would reduce the reliance on such distributional assumptions when forming designs \citep*{mcgree2023general,overstall2023gibbs}.
In addition, another potential limitation is the use of spline basis functions as more flexible modelling approaches may be needed in practice.  To address this, future work could consider alternatives such as Gaussian Processes \citep*{kennedy2001bayesian}, Neural Networks \citep*{beck2000improving} and Generalized Additive Models \citep*{wood2003thin} could be adopted.


\section{References}
\begingroup
\bibliographystyle{apalike}
\renewcommand{\section}[2]{}
\bibliography{bibliography.bib}
\endgroup

\newpage

\end{document}